# Probing static discharge of polymer surfaces with nanoscale resolution

Nikolay Borodinov[1,2], Anton V. Ievlev[1], Jan-Michael Carrillo[1], Andrea Calamari[3], Marc Mamak[3], John Mulcahy[3], Bobby G. Sumpter[1], Olga S. Ovchinnikova[1], Petro Maksymovych[1]

[1]Center for Nanophase Materials Sciences, Oak Ridge National Laboratory, Oak Ridge, Tennessee, United States

[2]Department of Materials Science and Engineering, Clemson University, Clemson, South Carolina, United States

[3]Research and Development, Procter & Gamble, Cincinnati, Ohio 45202, United States

**Abstract**

Triboelectric charging strongly affects the operation cycle and handling of materials and can be used to harvest mechanical energy through triboelectric nanogenerator set-up. Despite ubiquity of triboelectric effects, a lot of mechanisms surrounding the relevant phenomena remain to be understood. Continued progress will rely on the development of rapid and reliable methods to probe accumulation and dynamics of static charges. Here, we demonstrate *in-situ* quantification of tribological charging with nanoscale resolution, that is applicable to a wide range of dielectric systems. We apply this method to differentiate between strongly and weakly charging compositions of industrial grade polymers. The method highlights the complex phenomena of electrostatic discharge upon contact formation to pre-charged surfaces, and directly reveals the mobility of electrostatic charge on the surface. Systematic characterization of commercial polyethylene terephthalate samples revealed the compositions with the best antistatic properties and provided an estimate of characteristic charge density up to $5\times10^{-5}$ C/m$^2$. Large-scale molecular dynamics simulations were used to resolve atomistic level structural and dynamical details revealing enrichment of oxygen containing groups near the air-interface where electrostatic charges are likely to accumulate.

Triboelectric charging was historically the first demonstration of electricity known to humanity, and even after centuries of intense research and investigation this phenomenon is extremely relevant to the technologies of today. As materials of different nature come in contact, they may develop electrical charge which occurs universally in almost any industrial set-up. The resulting electrostatic attraction or repulsion hinders the handling of a very diverse pool of objects including pharmaceutical powders[1] and components of the integrated curcuits[2], which can be damaged by the electrostatic discharge. However, in some cases the effect of tribological charging may be utilized for the benefit of the practical application. One of the most prominent examples is the triboelectrostatic separation based on the variability of the sample response to contact electrification. This process is successfully utilized for mineral processing[3] and well as separation of granular plastics[4–6] and coal particles,[7] and size-dependent separations.[8] Ionic electrets[9] found their application in xerography printing, electret-based microphones and van der Graaf generators. Recently, numerous works addressed the prospects of triboelectric power nanogenerators to allow harvesting of mechanical energy.[10–12] Thus, the quantification of the triboelectric charge, visualization and quick identification of viable application-specific materials is needed to streamline the research and optimization of triboelectric systems and gain fundamental insight into triboelectric processes, particularly at length-scales comparable to charge localization and diffusion.

Scanning probe microscopy – a standard method to achieve high spatial resolution for functional imaging – has previously been used to investigate triboelectric processes. Several works reported nanoscale measurements of triboelectric charging on the surfaces of inorganic[13–15] and polymer[16–18] dielectrics. The common method in these cases was to employ probe scanning or probe-surface contact formation as a way to triboelecrically charge the surface with nanoscale resolution. While intriguing, one possible limitation in this case is the unknown rate of discharge in the contact region that may occur simultaneously with the charging process. This is particularly true if triboelectric charging is carried out by a semi-conducting probe, such as doped silicon. The other major difference is that the scanning rate of even the fastest atomic force microscopes is still far too slow compare to the characteristic displacement rate of macroscopic bodies in triboelectric processes (~1 cm/s). The net deposited charge in this case may therefore be significantly smaller than the intrinsic capacity of the material to sustain static charge. Moreover, the preferred method of detection of triboelectric charges has so far been Kelvin probe force microscopy, which necessitates the use of thin films, and limiting the applicability of these methods toward industrial grade materials.

Instead of focusing on localized charging, therefore, we developed a different approach that quantifies localized *discharge* of a previously charged surface. Furthermore, the underlying method to detect charges utilizes force-distance curves, a common measurement of interaction force between closely spaced bodies.[1,19–24] We demonstrate the capabilities of this methodology directly on industrial polymer samples – representing surfaces of commercial plastic bottles. Specifically, we investigate the efficiency of macroscopic charging, nanoscale discharging, and nanoscale recharging following a partial discharge. We directly observe the dynamics of the triboelectric charge and quantify its surface mobility, within the approximations of a continuum electrostatic model. At the same time, we show that it is possible to reliably differentiate polymer compositions based on their propensity to accumulate triboelectric charges. This is relevant to the handling of industrial bottles on production lines, where static charge can present significant challenges to handling and scale-up.

Polyethylene terephthalate studied here is one of the most important plastic used for packaging. In our case the surface of PET may be coated with mold-release agents, lubricants and anti-static additives. As these chemicals drift throughout the bulk polymer and across the surface, the properties of the interface are changing, which raises the question of the spatial distribution of these additives and their effect on the tribocharging.

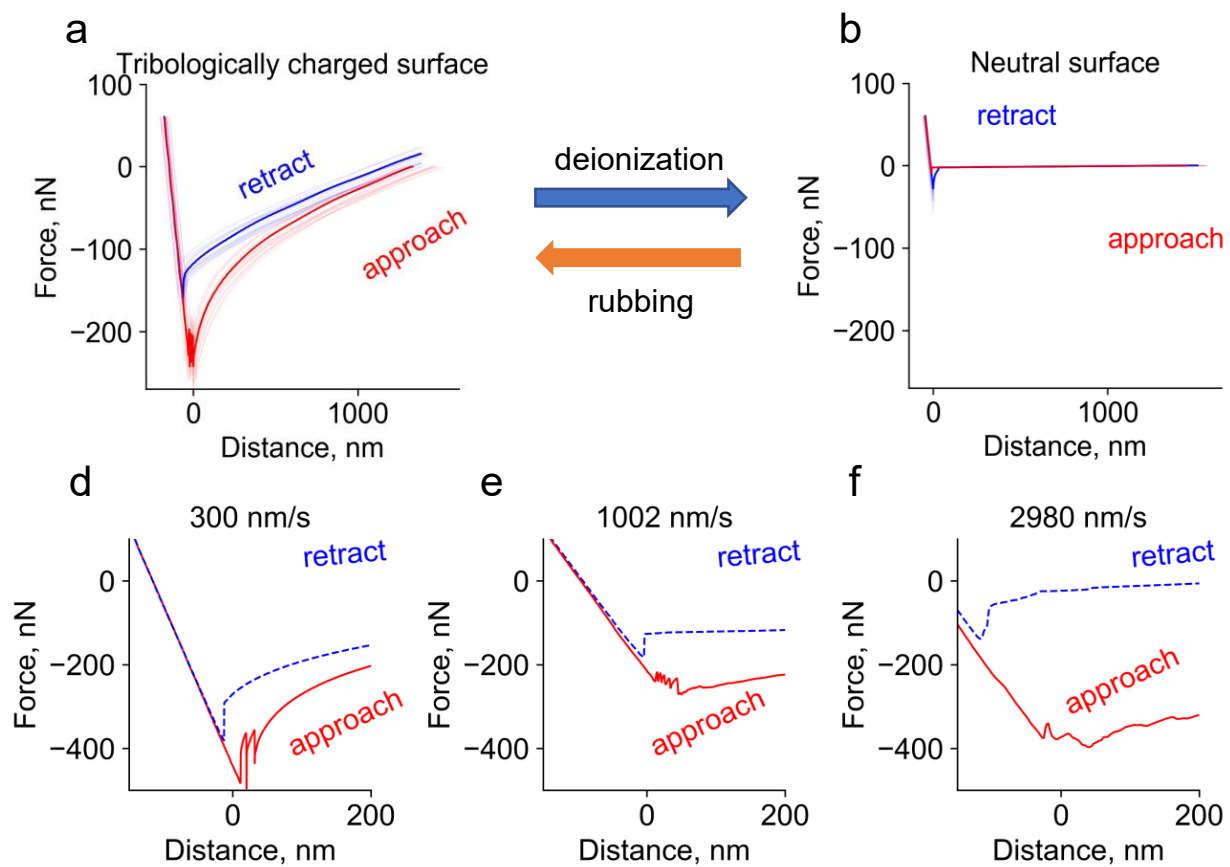

**Figure 1.** (a, b) Deionization/rubbing procedures can reversibly switch between charged (a) and neutral (b) polymer surfaces as evidenced by the large changes in contact and non-contact forces. (c, d, e) Discharge pattern in the contact region for different measurement rates: slower tip motion toward the surface leads to more well-defined sequential "jumps" in and out of contact with the surface.

Given the scatter of the properties of the as-made industrial bottles and their largely unknown history, including their partial charging, we first established a reliable procedure to charge the bottles over large areas. We utilized an electrically actuated nylon brush that was brought in contact with the bottle and brushed for a given period of time (approx. 10 min). A typical example of the force-distance curve (FDC) of the tribologically charged polymer is displayed in **Figure 1a**. Several quite pronounced features are observed, some notably contradicting the expectations for the force-distance curves on polymers: (1) the force at contact is very large, ~ 200-300 nN, about 10 times stronger than commonly reported values for PET; (2) there is a long-range slope on both approach and retract branches of the force-distance curve; (3) perhaps most notably, the forces upon approach are consistently stronger, than those upon retraction, following the formation and breaking of the mechanical contact with the polymer.

The long-range component of the force indicates a strong electrostatic force. To confirm this notion, we exposed the bottles to ionized air, neutralizing the built-in charge. Ionized air completely transforms the force distance curves, as seen in **Figure 1b**. The long-range component is removed. The difference between forces on approach and retraction becomes negligible, except for the contact region. Here the adhesive force is seen to be ~20-30 nN, in line with previous observations.[6] We also note, that electrostatic forces can be removed by sample washing with ethanol. Finally, the measurements of the electrostatic charge with the hand-held probe revealed that the in the case of nylon-PET contact the polyethylene terephthalate developed overall negative charge.

Once the PET polymer surface has been discharged, rubbing it with a nylon brush, brings back the charged state, essentially reverting the curve back to **Figure 1a**. This very reliable procedure allowed us to investigate the details of the triboelectric discharge in detail. As seen in **Figure 1a**, upon touching the surface, a grounded tip locally discharges the surface, reducing the attraction force. The structure of the force-distance curve in the contact region, however, is not trivial, Figure 1b. Upon approach, the tip is brought into contact with the surface when the electrostatic force overcomes the restoring force of the AFM cantilever. Instead of staying in contact, however, the tip then abruptly separates from the surface of the polymer. A short instance after, the tip again jumps into contact, and again separates from it. This cycle repeats several times before the final contact is made, and the force-distance curve transitions into an elastic region where the tip is held in contact with te surface.

The origin of the repeated "jumps" can be understood by considering the decreasing value of the force the tip experiences after intermittent contact. It is clear from **Figure 1c-e** that the value of the non-contact force progressively decreases, toward its "saturation" value established after prolonged contact with the surface. Therefore, each of the intermittent contact event partially discharges the region of the immediate contact. The repetition of the cycles then requires that the charge density at least partially restores in the contact region, so that to attract the tip into contact again. This kinetic picture, where the triboelectric charge has finite mobility, is supported by the dependence of the "jumps" on the approach speed of the tip: the jumps become substantially less resolved at higher approach speed (**Figure 1c-e**), suggesting a characteristic time-scale for the recharge cycle for the small near-contact region is ~ 100 ms.

Using finite element modeling (see supporting information) the amount of the charge lost at each jump was estimated to reach 5-7%. **Figure S1b** (supporting information **section 1**) displays the fitting of the electrostatic part of an exemplary force-distance curve using the COMSOL model described above. As evident, the modeling satisfactorily predicts the non-contact region of the force-distance curve and allows for the estimation of the surface charge. The force acting on a tip located 1 nm away from the contact scales with the charge quadratically, allowing us to estimate the charge density to be approximately $2\times10^{-5}$ C/m$^2$. For comparison, the measurement of the surface charge and mobilities done by SKPM (scanning Kelvin probe microscopy) for the tribological charging of silica revealed[15] charge density to reach $5*10^{-5}$ C/m$^2$.

In order to make the analysis of the force-distance curves more systematic, we introduced a set of five empirical parameters: (1) maximum contact force (*maxforce*) – directly measured in the force-distance curve; (2) *pull-off* force related to the adhesion between the tip and PET; (3) *hysteresis*, which is the difference in force extrapolated into the points of contact, (4) *surge*, which is the extrapolated first derivative of the approach curve in the point of contact and (5) *jounce*, which is second derivative of the approach curve in the same point, as shown in **Figure 2a** (see supporting information **section 2** for detailed definition of these parameters). Hereafter we will use these parameters to quantify the evolution of individual force-distance curves, and help us classify different polymer compositions. Specifically, the maximum force parameter appears to be an effective descriptor of the charge density under the tip.

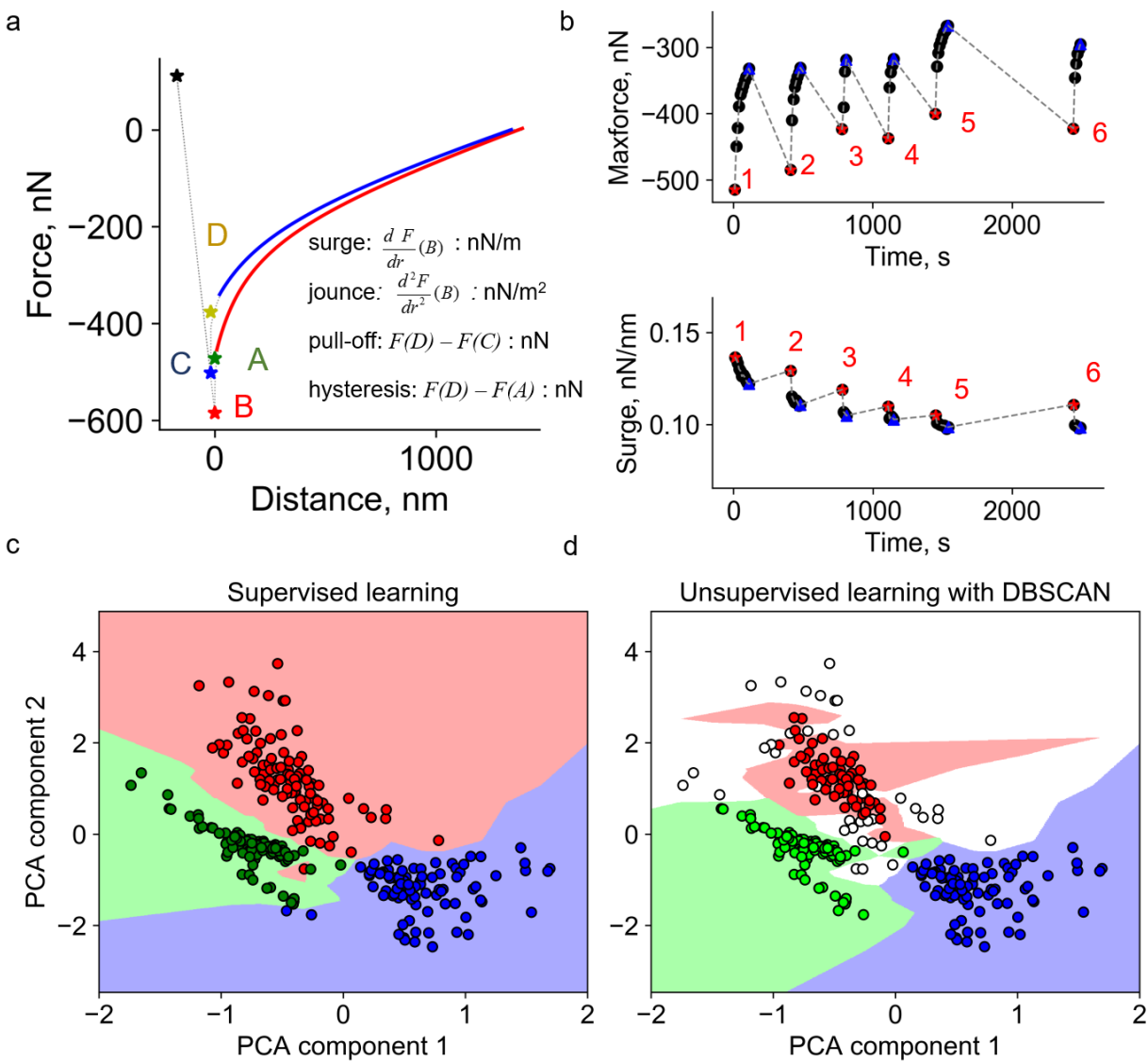

**Figure 2**. The parametrization of the electrostatic curve (a), the migration of the electric charges as a result of the interrogation of a single spot: after the first contact (red star) the localized charge is decreased until the last contact in the series (blue triangle), however, after a period of time the charge is restored which is reflected by the corresponding parameters, (c) the decision boundaries drawn using supervised learning done by nearest neighbors classification, colors reflect one of three commercial PET compositions, (d) decision boundaries drawn using the DBSCAN clustering algorithm, the color reflects the membership of the same cluster (white dots are points that are not assigned to any cluster, outliers).

To probe the time-scale of charge-discharge events more accurately, we repeatedly interrogated the single spot of a pre-charged surface of a polymer with variable timing between individual force distance curves. The initial contact (denoted by a red star #1 **in Figure 2b**, see supporting information **section 3**) drains the charge which after several indentations stops changing (denoted by a blue dot). After approximately 5 minutes where the tip is not in contact, the drift of the charges across the surface partially restores the original state and force- distance curves display strong electrostatics again (red stars 2 through 6 in **Figure 2a**). The fast-discharge/slow-recharge cycle can subsequently be repeated as seen in **Figure 2b**. In order to roughly estimate the diffusion coefficient, we can consider Brownian motion of charge on the surface so that[15]:

$$\frac{<l>^2}{2} = D\Delta t \qquad (1)$$

Where, is <l> the time-averaged displacement of charge, $D$ – diffusion coefficient and $\Delta t$ – time between observation. If we assume 1 micrometer range of the charge migration and $\Delta t = 300$ s, the diffusion coefficient is about $10^{-15}$ m$^2$/s.. Mobility in the range of $10^{-15}$-$10^{-16}$ m$^2$/s was previously reported for $SiO_2$.[15]

When a larger area (20x20 µm$^2$) is repeatedly probed over a grid of force distance curves with ~2 µm resolution, eventually the surface becomes stripped off its charge. This gradual discharge of the entire area is highlighted in **Figure 3a** which displays parameters *maxforce* and *surge* for a series of grids. Both parameters are gradually reduced signaling the discharge of the whole surface area. This effect is also seen in the empirical parameter space in **Figure 3b**, where the centroids of the corresponding values for each grid is shifting toward progressively smaller values. In addition to the overall discharge process, certain points on the map reveal further evidence of charge migration. The most notable of this is the "stripe" pattern, where (see also supporting information **section 4**). The stripes correspond to slight alternation of the parameters of the force-distance curve along the slow direction of the grid (top to bottom in **Figure 3a**). As the tip harvests the charge on a specific line, the neighboring line is also partially affected due to finite mobility of charge. Strictly speaking, for the stripes to occur the discharge kinetics should be non-linear in time, which will be investigated in the future. However, the "stripes" also necessitate mobility on a micrometer length scale, consistent with our prior estimate. Finally, the grids also show that in order to ensure the complete elimination of the triboelectric attraction of the specimens in the industrial applications, the point contact or localized discharge will be ineffective, as the charges presented in the vicinity would nullify the discharge effect.

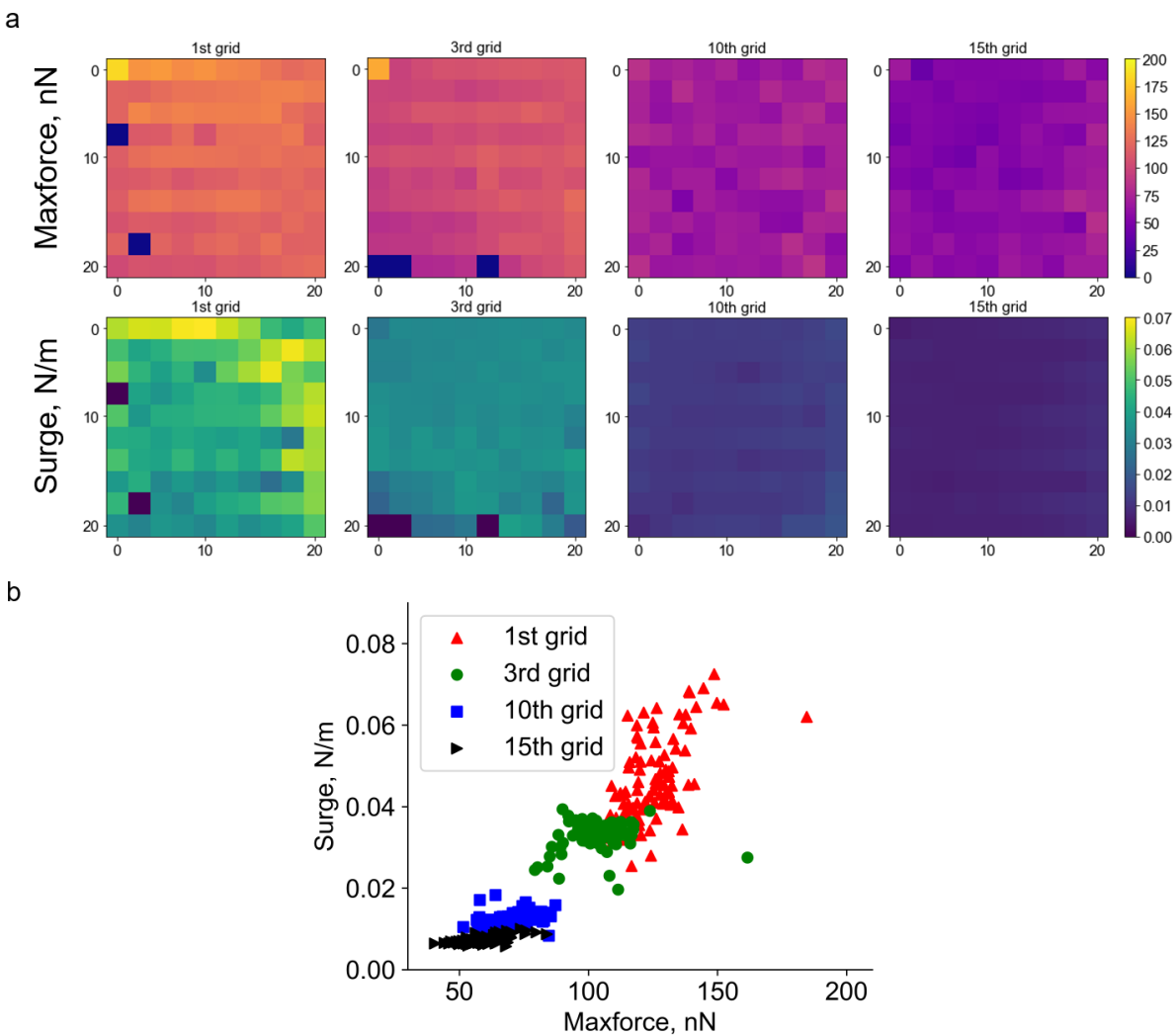

**Figure 3.** (a) Parameters *maxforce* and *surge* for the force-distance curves obtained on a grid of points across the PET surface. The overall discharge is rather clearly. The size of the grid is 20x20 µm$^2$, the points where the fitting procedure failed amount to no more than 3% of the dataset. They are excluded from the statistical analysis and marked as "0". (b) Surge plotted as a function of *maxforce* for several grid datasets.

The reproducibility of the charge-discharge process also allows for quantitative differentiation between specific polymer compositions. To this end, we have sampled 100-1000 FDC curves for three compositions of PET with different additives supplied by Procter & Gamble Co. The exemplary histograms of the parameters are presented in supporting information **section 5**. It is clearly seen that it is very easy to distinguish between charged and neutral states of the dielectric PET films. Moreover, there do appear to be distinct differences between individual polymer compositions, signifying the differences in either their charging ability or specific interaction with the nylon brush.

To reduce the dimensionality of the parameter space spanned by the five empirical parameters described above, we applied Principal Component Analysis and retained two most significant components to represent the whole space. The projections of individual parameters for each force-distance curve and polymer composition are shown in **Figure 3c**. It is evident that the compositions can indeed be well-separated. Using statistical classification methods, the observed projections can further be converted into a classifier, to identify the regions of parameter space corresponding to a specific composition. **Figure 2c,d** compares the performance of the k-nearest neighbor (KNN) clustering with supervised learning (where the compositions are explicitly labeled, **Figure 2c**) and unsupervised learning, where the parameters are automatically clustered using the DBSCAN algorithm (**Figure 2d**).[25] Both approaches perform reasonably well, separating the cluster centers. However, arguably the unsupervised learning performs better, because DBSCAN efficiently identifies data at the decision boundaries as outliers, whereas the supervised clustering creates a spurious region (green) that is serparated from the wide area corresponding to the first composition. Thus, our methodology reveals that distinct PET compositions demonstrate different ability to accumulate tribological charge and can be discerned based on this quality.

Finally, it is interesting to ask where the triboelectric charge can be localized on the PET, so as to maintain the observed mobility. To understand the morphology of the PET near the air-interface, we used large-scale atomistic molecular dynamics simulations to show: (1) in a mixture of long and short PET polymer chains, short chains are more abundant at the air-polymer interface, and (2) it is more probable to locate polymer chain-ends at the air-polymer interface. The synthesis of commercial PET involves two steps, where the second step is a linear polycondensation process, for which the degree-of-polymerization (DP) is increased. The theoretical limit for the polydispersity index (PDI) is 2 for polycondensation (or PDI=1+$p$, where $p$ is conversion and ~1) and the high value of the PDI suggests that long chains are mixed with shorter chains. In **Figure 4a**, the molar density of oxygen for our model system that contains two types of hydroxyl terminated PET chains, is shown. The MD simulation box contains equal number of PET chains having DP's of 10 and 3 at a temperature of 300 ˚C or above the melting point of PET. The simulation box is configured as a slab where $z=0$ Å is the location of the middle of the slab. The $z$-dependent local molar densities, $\rho(z)$, in **Figure 4** is normalized by the molar density at the middle of the slab, $\rho$, to compare the relative amount of oxygen at the middle of the slab against those found at the interface. Clearly, the oxygen from DP=3 are relatively more abundant near the interface in comparison to the oxygen from DP=10.

Next, we performed simulations for a box containing PET homopolymers with DP=10. We performed a similar analysis, but this time we tracked the location of the hydroxyl oxygen found at the chain ends. This was done in order to locate where the chain ends are positioned relative to the middle of the slab. The hydroxyl oxygen serves as a pointer tagging the location of nearby oxygen found in the ester linkages. **Figure 4b** shows that the relative amount of hydroxyl oxygen is higher at the air-interface in comparison to the total molar density, indicating that it is more probable to find hydroxyl oxygens at the interface rather than at the middle of the slab.

In summary, the MD simulations suggest that the probable molecular moieties involved in the electrostatic charge accumulation are the oxygen in the shorter and more mobile PET chains, and with greater contributions from oxygen in the ester linkages near the chain ends as well as the oxygen at the chain ends. This was corroborated by taking the configurations of the PET chains obtained from the MD simulations and computing where extra charge would coalesce using semiempirical quantum chemistry (PM3) calculations. Surface localization of charge is qualitatively consistent with our ability to discharge the surfaces, and the observed charge mobility. The difference between polymer compositions observed in **Figure 3** can stem from a variety of factors, but it seems plausible to suggest that the density and configuration of the electroactive groups on the surface will be the primary factor.

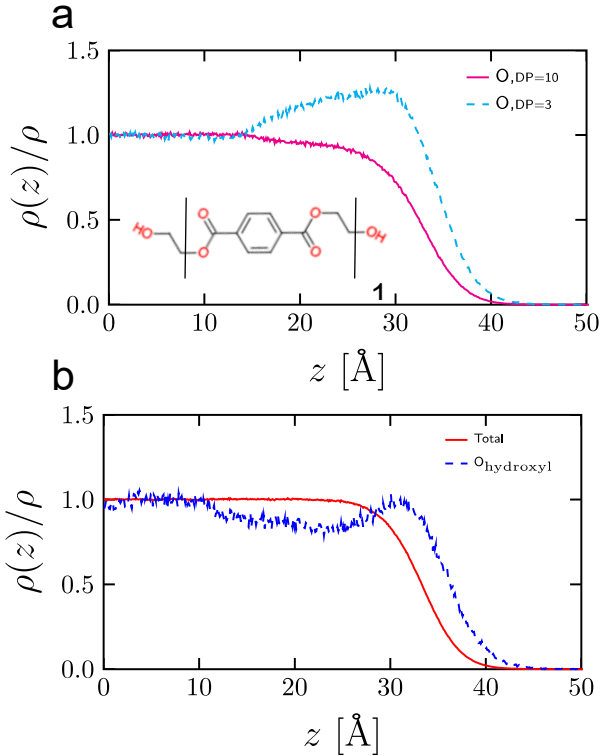

**Figure 4.** (a) Normalized local molar density, $\rho(z)/\rho$ of oxygen atoms from an equimolar mixture of PET chains with DP=3 (cyan) and DP=1 (magenta). (b) Normalized local molar density, $\rho(z)/\rho$ from a homopolymer melt of PET chains with DP=10 of the total molar density (red) and oxygen atoms at the chain ends (blue). The inset in (a) shows the monomer unit of the PET model, which is terminated with hydroxyl groups, used in the atomistic molecular dynamics simulation of a slab where the middle of the slab is at $z$=0 Å and the air-interface is at $z$~45 Å.

**Conclusion.** The quantification of the tribological charge on an industrial-grade samples has been performed using the analysis of force-distance curves within an atomic force microscopy set-up. This technique probes the rate and efficiency of *discharge* of a charged polymer in the nanoscale contact to a metallic asperity. The issues of surface charge migration and complex phenomena of repeated contact within the single approach curves are considered in detail and are quantified using the modelling. The parameters extracted from the force-distance curves were used to estimate the density of the triboelectric charge, detect and estimate the mobility of electrostatic charge on the surface and to compare the performance of the polyethylene terephthalate compositions. We expect that this technique can be successfully used for the quick profiling of the materials for the desired charge accumulation characteristics. For example, the method of force-distance curve analysis can in principle be used for the estimation of the most promising candidates for the triboelectric nanogenerators or selection of the antistatic coating.

**Acknowledgement**

This research was carried out at the Center for Nanophase Materials Sciences, a US Department of Energy Office of Science User Facility. The scope of the work was under a CRADA between Proctor & Gamble Co. and Oak Ridge National Laboratory. This research also used resources of the Oak Ridge Leadership Computing Facility at the Oak Ridge National Laboratory, which is supported by the Office of Science of the U.S. Department of Energy under Contract No. DE-AC05-00OR22725.